\documentclass[prb,twocolumn,superscriptaddress,showpacs]{revtex4}
\usepackage{graphicx}
\usepackage{dcolumn}
\usepackage{bm}

\begin{document}

\title{Spin precession in anisotropic media}


\author{B. Raes}
\affiliation{Catalan Institute of Nanoscience and Nanotechnology (ICN2), CSIC and The Barcelona Institute of Science and Technology, Campus UAB, Bellaterra, 08193 Barcelona, Spain}
\author{A. W. Cummings}
\affiliation{Catalan Institute of Nanoscience and Nanotechnology (ICN2), CSIC and The Barcelona Institute of Science and Technology, Campus UAB, Bellaterra, 08193 Barcelona, Spain}
\author{F. Bonell}
\affiliation{Catalan Institute of Nanoscience and Nanotechnology (ICN2), CSIC and The Barcelona Institute of Science and Technology, Campus UAB, Bellaterra, 08193 Barcelona, Spain}
\author{M. V. Costache}
\affiliation{Catalan Institute of Nanoscience and Nanotechnology (ICN2), CSIC and The Barcelona Institute of Science and Technology, Campus UAB, Bellaterra, 08193 Barcelona, Spain}
\author{J. F. Sierra}
\affiliation{Catalan Institute of Nanoscience and Nanotechnology (ICN2), CSIC and The Barcelona Institute of Science and Technology, Campus UAB, Bellaterra, 08193 Barcelona, Spain}
\author{S. Roche}
\affiliation{Catalan Institute of Nanoscience and Nanotechnology (ICN2), CSIC and The Barcelona Institute of Science and Technology, Campus UAB, Bellaterra, 08193 Barcelona, Spain}
\affiliation{Instituci\'{o} Catalana de Recerca i Estudis Avan\c{c}ats (ICREA), 08070 Barcelona, Spain}
\author{S. O. Valenzuela$^{*}$}
\affiliation{Catalan Institute of Nanoscience and Nanotechnology (ICN2), CSIC and The Barcelona Institute of Science and Technology, Campus UAB, Bellaterra, 08193 Barcelona, Spain}
\affiliation{Instituci\'{o} Catalana de Recerca i Estudis Avan\c{c}ats (ICREA), 08070 Barcelona, Spain}

\date{\today}

\begin{abstract}
We generalize the diffusive model for spin injection and detection in nonlocal spin structures to account for spin precession under an applied magnetic field in an anisotropic medium, for which the spin lifetime is not unique and depends on the spin orientation. We demonstrate that the spin precession (Hanle) lineshape is strongly dependent on the degree of anisotropy and on the orientation of the magnetic field. In particular, we show that the anisotropy of the spin lifetime can be extracted from the measured spin signal, after dephasing in an oblique magnetic field, by using an analytical formula with a single fitting parameter. Alternatively, after identifying the fingerprints associated with the anisotropy, we propose a simple scaling of the Hanle lineshapes at specific magnetic field orientations that results in a universal curve only in the isotropic case. The deviation from the universal curve can be used as a complementary means of quantifying the anisotropy by direct comparison with the solution of our generalized model. Finally, we apply our model to graphene devices and find that the spin relaxation for graphene on silicon oxide is isotropic within our experimental resolution.
\end{abstract}


\maketitle

\section{Introduction}

Two-dimensional materials (2DMs), such as graphene, phosphorene and transition metal dichalcogenides, are gathering increasing attention from the spintronics community because of the tunability of their transport properties \cite{geimNM,castronetoRMP,phosphorene,reviewTMDs,NatNanotechnol.9.324.2014,JPhysD.47.094011.2014,2DMaterials.2.030202.2015,NatPhys.8.563.2013,NatPhys.11.830.2015,gate}. The reduction of the thickness of the device channel down to the atomic scale results in enhanced gate control and opens the door for subtle material engineering \cite{geim2DHetero}, where properties such as magnetism or large spin-orbit coupling (SOC) could be borrowed from other materials in proximity with the 2DM \cite{2DMaterials.2.030202.2015,PhysRevB.93.241401.2016}. However, as their thickness is reduced, 2DMs become increasingly susceptible to environmental effects that modify the spin dynamics. In addition, the 2D character of their structure make them highly anisotropic, which is reflected in their electronic, optical and mechanical response. Spin transport can also be anisotropic if, for example, the spin dynamics is governed by spin-orbit fields (SOFs) with a dominant orientation \cite{APSlov.57.565.2007}. A classical example is the two-dimensional electron gas (2DEG) with Rashba SOC, for which the lifetime of spins oriented in the plane of the 2DEG, $\tau_{s\parallel}$, is twice as large as the lifetime of spins oriented perpendicular to it, $\tau_{s\perp}$; that is $\zeta\equiv\tau_{s\perp}/\tau_{s\parallel} =0.5$. This is due to the fact that spins along the in-plane Rashba field do not precess and do not dephase. The spin relaxation anisotropy, quantified by the anisotropy ratio $\zeta$, is therefore a parameter that is particularly relevant for characterizing the spin dynamics in an anisotropic system. Indeed, it has been pointed out that it could be crucial to identify the microscopic mechanisms at play in the spin relaxation in graphene \cite{NatNanotechnol.9.324.2014}. However, the first attempts to measure it in transport experiments required large magnetic fields, which make the results only reliable at high carrier densities due to magnetoresistive effects \cite{PhysRevLett.101.046601.2008}.

\begin{figure}[h]
\includegraphics[width=0.7\linewidth]{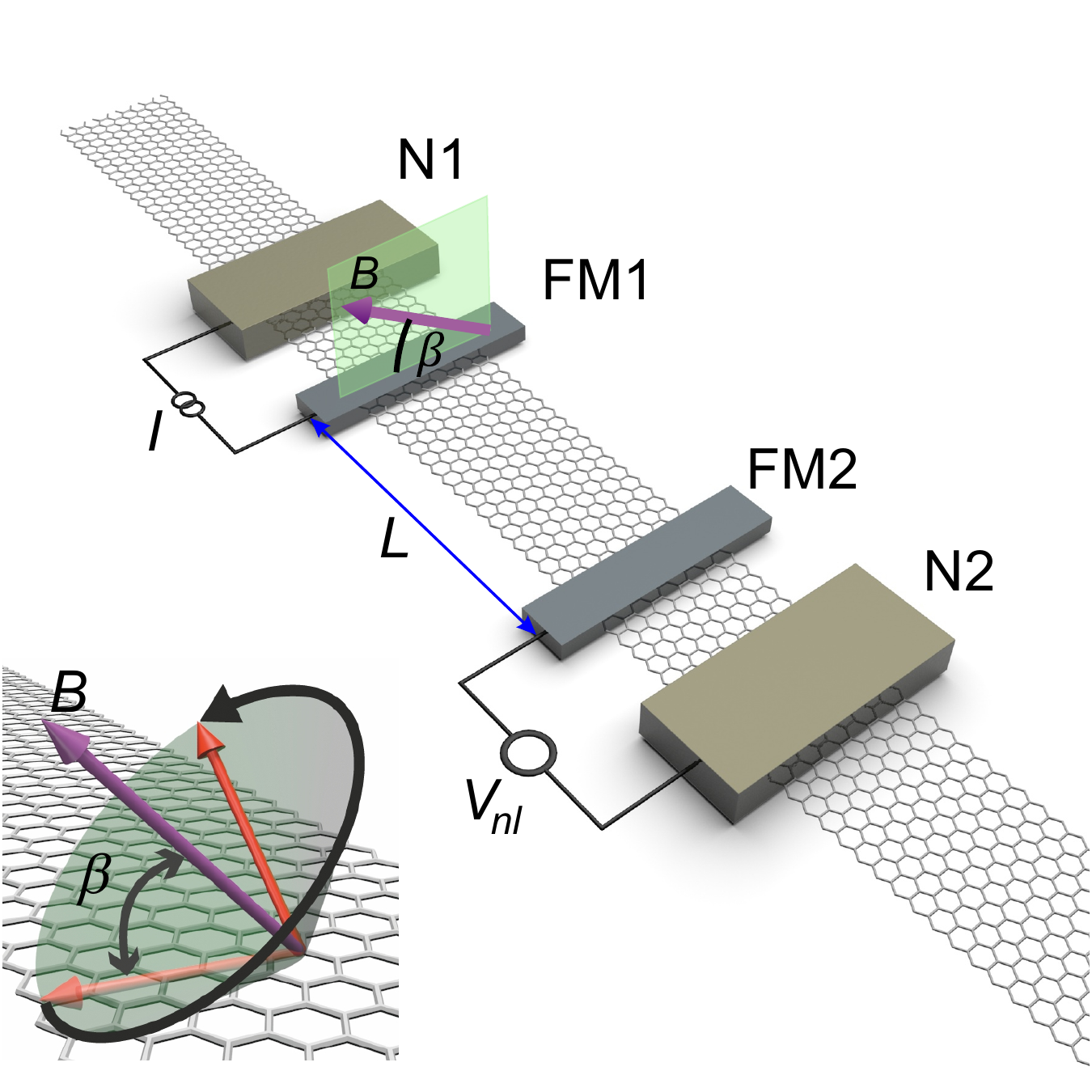}
\caption{Schematic drawing of the lateral spin device geometry showing both the outer normal metallic electrodes (N1 and N2) and the inner ferromagnetic injector (FM1) and detector (FM2) electrodes. Wiring is shown in the nonlocal configuration, in which a current $I$ is injected between FM1 and N1 and the nonlocal voltage $V_\mathrm{nl}$ is measured between FM2 and N2. Inset:  Schematic illustration of the oblique spin precession configuration investigated in this article. The magnetic field $\vec{B}$ is applied in a plane that contains the easy axis of the ferromagnetic electrodes and that is perpendicular to the substrate (see main panel and Fig. \ref{Fig1b}). For an oblique field, that is $\beta\neq0,90^\circ$, the spins precess out-of-plane as they diffuse from FM1 towards FM2. In this situation, we demonstrate that the effective spin lifetime is sensitive to both parallel and perpendicular spin lifetimes, $\tau_\mathrm{s\parallel}$ and $\tau_\mathrm{s\perp}$, and the spin relaxation anisotropy can be experimentally obtained.}\label{Fig1a}
\end{figure}

We have recently demonstrated that $\zeta$ can be readily determined by measuring the response of nonlocal spin devices under oblique magnetic fields \cite{BR2016}. The magnetic field $\vec{B}$ is applied in a plane whose components are perpendicular to the substrate and parallel to the easy axis of the ferromagnetic electrodes, which act as spin injector and detector. The angle $\beta$ characterizes the field orientation  (see Fig. \ref{Fig1a}).
The essential idea is to investigate the non-precessing spin component along the applied magnetic field, $s_{B\parallel}$, for different $\beta$. One can determine whether $\zeta<1$, $\zeta=1$  or $\zeta>1$ by observing if the spin relaxation of $s_{B\parallel}$ for $\beta\neq0, 90^{\circ}$ is faster, equal or slower, respectively, than the spin relaxation in the graphene plane.
In Ref. \onlinecite{BR2016}, we demonstrated that the effective relaxation time of $s_{B\parallel}$, $\tau_{s B}$, follows a simple relationship with $\zeta$ and $\beta$. Using this relationship, one can obtain $\zeta$ by fitting the nonlocal response versus $\beta$, where $\zeta$ is the only fitting parameter.

However, the full characterization of spin transport is typically achieved by performing spin precession (Hanle) experiments, \cite{js1985,SOV2009} as demonstrated in a variety of materials in the diffusive limit, including metals \cite{js1985,Nature.416.713.2002}, semiconductors \cite{Nature447.295.2007,NatPhys3.202.2007} and graphene \cite{Nature448.571.2007}. While diffusing towards the detector electrode, the injected spins undergo Larmor precession about $\vec{B}$, which in the standard configuration is oriented perpendicular to the channel material, $\beta=90^{\circ}$. Changing the field strength alters the precession angle $\phi$ at the detector electrode and results in a modulation of the detected nonlocal spin signal. The spatio-temporal evolution of the spins is commonly described by diffusive Bloch equations. The solution of these equations in the isotropic limit have enabled the evaluation of fundamental spin transport parameters from direct fitting of the magnetic-field-dependent nonlocal signals. When the magnetic field is perpendicular to the channel, the spins stay in the plane and, even if the material presents a perpendicular anisotropy, the spin relaxation is determined by  $\tau_{s\parallel}$ and no signature of the anisotropy is observed in the Hanle lineshape. In oblique magnetic fields, such as those in Ref. \onlinecite{BR2016}, the spins precess out of plane and the spin dynamics becomes sensitive to $\tau_{s\perp}$. Under these conditions, it is reasonable to expect that the experimental lineshape of a Hanle measurement will manifest signatures of the anisotropy. The analysis of the spin precession lineshape is thus complementary to investigating the relaxation of the non-precessing spin component along $\vec{B}$, which we have developed in Ref. \onlinecite{BR2016} and, therefore, it is of high interest for spin relaxation anisotropy studies.

In this paper, we solve the anisotropic Bloch equations and discuss the evolution of the expected nonlocal signal as a function of $\zeta$ and $\beta$ in order to identify the anisotropy fingerprints in the oblique spin precession experiments. We find that both the magnitude and position of the local extremum for the collective $\pi$ spin precession strongly depend on $\zeta$. After a suitable scaling the spin precession lineshape for $\zeta=1$ is independent of $\beta$, which is not the case for $\zeta\neq 1$. We thus propose a simple procedure to determine the anisotropy based on this scaling behavior, and then implement it experimentally using graphene nonlocal spin devices. We show that the anisotropy can be characterized with only two magnetic field orientations without knowing the orientation angles precisely, which could be advantageous in some experimental setups.

The paper is organized as follows. In Section \ref{Bloch}, we introduce a suitable device geometry to determine the spin-relaxation anisotropy and derive the diffusive anisotropic Bloch equations to describe it. In Sections \ref{iso} and \ref{aniso}, we solve the equations derived in Section \ref{Bloch} in the isotropic and anisotropic limits, respectively. We show that the spin precession under oblique magnetic fields presents very specific fingerprints that give access to the anisotropy of the system under study. In Section \ref{aniso}, we also discuss the influence on the measured anisotropy that may result from the presence of the contacting electrodes. In Section \ref{Exp}, we present experimental results on the spin relaxation anisotropy in graphene, and conclusions are given in Section \ref{Concl}.


\section{Bloch equations for an anisotropic system} \label{Bloch}
Spin precession experiments in a spin transport channel are typically performed in a nonlocal lateral geometry using a four-terminal device having two inner ferromagnetic (FM) electrodes and two outer electrodes, which are ideally nonmagnetic. Such a device is shown schematically in Fig. \ref{Fig1a}. As introduced above, in standard spin precession measurements the magnetic field $\vec{B}$ is applied perpendicularly to the substrate on which the device is fabricated ($\beta=90^{\circ}$). In this configuration, $\vec{B}$ is also perpendicular to the easy magnetization axis of the FM electrodes, which, following the geometrical ferromagnetic anisotropy, is parallel to their length. As the spins diffuse from the injector (FM1) to the detector (FM2), the spin precession around $\vec{B}$ proceeds in the plane of the channel and thus only the relaxation of spins having an in-plane orientation is involved. If the spin relaxation in the channel is isotropic, such an experiment provides all of the information that is required to fully characterize the spin dynamics in the material. In general, however, spin transport can be anisotropic, where spin lifetimes $\tau_{s\parallel}$ and $\tau_{s\perp}$ for spin orientations parallel ($\parallel$) and perpendicular ($\perp$) to the channel plane are different \cite{Note0}. The anisotropy, which can result from spin-orbit fields with a preferential orientation, can be  characterized by the ratio between these spin lifetimes $\zeta\equiv\tau_{s\perp}/\tau_{s\parallel}$. For SOFs pointing preferentially in the channel plane, we expect $\zeta<1$, while for SOFs perpendicular to the plane, we expect $\zeta>1$. However, if the main relaxation mechanism is driven by random magnetic impurities or SOFs with no preferential orientation, no anisotropy is expected \cite{NatNanotechnol.9.324.2014} and $\zeta=1$.

As argued in Ref. \onlinecite{BR2016}, spin precession experiments around an oblique magnetic field \cite{PhysRevB.68.245319.2003,ianOblique} overcome the limitation of standard measurements and allow the study of the spin lifetime anisotropy, which is of critical importance to determine the nature and preferential direction of the SOFs. The fundamental idea is simple: the precessional motion of the injected spins around the oblique magnetic field induces spin components perpendicular to the channel plane (see Fig. \ref{Fig1a}, inset) and as such the spatio-temporal evolution of the spin density, $\vec{s}(x,t)$, is sensitive to both $\tau_{s\parallel}$ and $\tau_{s\perp}$.

For an isotropic spin transport medium, $\tau_{s\parallel}=\tau_{s\perp}$, the Bloch equations can be solved analytically with suitable boundary conditions. In this case, the measured field dependence of the nonlocal signal can be described with a closed analytical expression. This model has been successfully used in practical modeling of spin-related phenomena in metals, semiconductors and graphene in the standard spin precession configuration \cite{js1985,Nature.416.713.2002,Nature447.295.2007,NatPhys3.202.2007,Nature448.571.2007,SOV2009}.

\begin{figure}[t]
\includegraphics[width=0.7\linewidth]{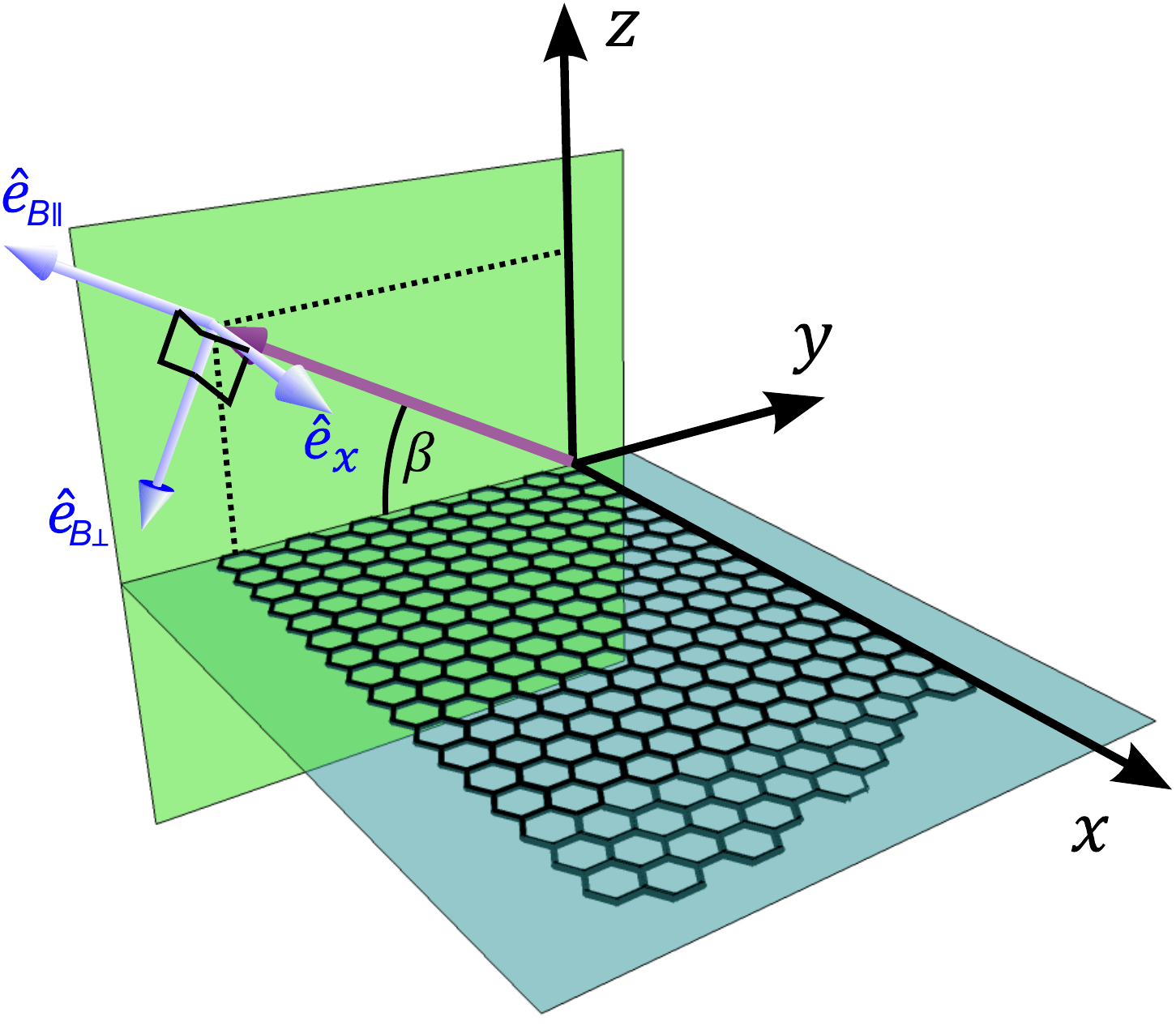}
\caption{Cartesian axes used for the calculation of the spatial evolution of the spin density. The unit vector $\hat{\mathbf{e}}_x$ is along the spin propagation channel $x$, while $\hat{\mathbf{e}}_{B_\parallel}$ and $\hat{\mathbf{e}}_{B_\perp}$ are along the parallel and perpendicular directions relative to the magnetic field $\vec{B}$, respectively.}\label{Fig1b}
\end{figure}

For an anisotropic spin transport medium, we extend the model to include $\tau_{s\parallel}$ and $\tau_{s\perp}$ explicitly, with $\zeta$ not necessarily equal to unity, and with arbitrary $\beta$. Within a rotated cartesian axis system determined by the unit vectors ($\hat{e}_x$,$\hat{e}_{B_\parallel}$,$\hat{e}_{B_\perp}$), as shown in the inset of Fig. \ref{Fig1b}, the diffusive Bloch equations are given by
\begin{eqnarray}\label{Eq1}
\frac{\partial \vec{s}}{\partial t}=\overline{D_s}\nabla^2\vec{s}+\gamma_c \vec{s} \times \vec{B}-\overline{\tau_{s}^{-1}}\cdot \vec{s},
\end{eqnarray}
with $\vec{s}(x,t)=(s_x,s_{B_\parallel},s_{B_\perp})$, $\vec{B}=(0,B,0)$ and
\begin{eqnarray}\label{Eq2}
\overline{\tau_{s}^{-1}}=\left(
                            \begin{array}{ccc}
                              \langle\tau_{xx}\rangle ^{-1} & \langle\tau_{x{B_\parallel}}\rangle ^{-1} & \langle\tau_{xB_\perp}\rangle ^{-1} \\
                              \langle\tau_{{B_\parallel}x}\rangle ^{-1} &  \langle\tau_{{B_\parallel}{B_\parallel}}\rangle ^{-1}  & \langle\tau_{{B_\parallel}B_\perp}\rangle ^{-1} \\
                              \langle\tau_{B_\perp x}\rangle ^{-1} & \langle\tau_{B_\perp{B_\parallel}}\rangle ^{-1} &  \langle\tau_{B_\perp B_\perp}\rangle ^{-1} \\
                            \end{array}
                          \right)
\end{eqnarray}
a symmetric matrix with
\begin{eqnarray}\label{Eq3}
\langle\tau_{{B_\parallel}{B_\parallel}}\rangle ^{-1}=\frac{1}{\tau_{s\parallel}\tau_{s\perp}}(\tau_{s\perp}\cos^2(\beta)+\tau_{s\parallel}\sin^2(\beta)), \\ \nonumber
\langle\tau_{B_\perp B_\perp}\rangle ^{-1}=\frac{1}{\tau_{s\parallel}\tau_{s\perp}}(\tau_{s\perp}\sin^2(\beta)+\tau_{s\parallel}\cos^2(\beta)), \\ \nonumber
\langle\tau_{B_\parallel B_\perp}\rangle ^{-1}=\langle\tau_{B_\perp B_\parallel }\rangle ^{-1}=\frac{ (\tau_{s\parallel}- \tau_{s\perp})}{\tau_{s\parallel}\tau_{s\perp}}\cos(\beta)\sin(\beta),\\ \nonumber
\langle\tau_{xB_\parallel}\rangle ^{-1},\langle\tau_{xB_\perp}\rangle ^{-1}=0\hspace{2mm} \mathrm{and}\hspace{2mm}\langle\tau_{xx}\rangle ^{-1}=\frac{1}{\tau_{s\parallel}}.
\end{eqnarray}

In the first term on the right-hand side of Eq. (\ref{Eq1}), $\overline{D_s}$ is considered to be a scalar matrix with all diagonal entries equal to $D_s$, the spin diffusion constant, which in general is different from the charge diffusion constant. The second term represents the torque, $\vec{N}=\gamma_c \vec{s} \times \vec{B}$, that $\vec{B}$ exerts on the spins, and that drives the precessional evolution of the spin density. The constant pre-factor, $\gamma_c=g\mu_B/\hbar$, is the gyromagnetic ratio of the carriers, where $\mu_B$ is the Bohr magneton and $g$ is the $g$-factor. In general $g$ depends on the material of interest and can be anisotropic \cite{gfactor}, although the anisotropic component is typically two order of magnitude smaller than the isotropic part \cite{gfactor} and can be disregarded. The last term introduces the spin relaxation, where $\overline{\tau_{s}^{-1}}$ is a ($3 \times3$) matrix with entries that are determined by $\zeta$ and $\beta$, as shown in Eqs. (\ref{Eq2}) and (\ref{Eq3}).


\section{Isotropic spin precession with arbitrary $\beta$}\label{iso}

For isotropic spin transport, the spin relaxation time matrix, Eq. (\ref{Eq2}), becomes diagonal. We solve the set of equations (\ref{Eq1})-(\ref{Eq3}) for the case of arbitrary $\beta$. In a nonlocal spin device, the voltage probes the projection of $\vec{s}=(s_x,s_\parallel,s_\perp)$ at position $x=L$ along the magnetization of FM2. If we consider that the magnetization orientation of the injector and detector remain fixed along their easy magnetization axis, the nonlocal voltage, normalized to the value at zero magnetic field, $V_{nl}(B,\beta)$, is given by
\begin{eqnarray}\label{Eq4}
&V_{nl}(B,\beta)=\cos^2(\beta)+ \sin^2(\beta)  V_{nl}(B_\perp)
\end{eqnarray}
where $V_{nl}(B_\perp)$ is the normalized nonlocal voltage when $\vec{B}$ is applied perpendicular to the channel ($\beta=90^\circ$),
\begin{widetext}
\begin{eqnarray}\label{Eq5}
&V_{nl}(B_\perp)= \sqrt{\frac{1}{2}}\frac{1}{f(b)}\Big[\sqrt{1+f(b)}\cos\Big(\frac{\mid b \mid}{\sqrt{1+f(b)}}\sqrt{\frac{L^2}{2\tau_{s}D}}\Big)-\frac{\mid b\mid}{\sqrt{1+f(b)}} \sin\Big(\frac{\mid b \mid}{\sqrt{1+f(b)}}\sqrt{\frac{L^2}{2\tau_{s\parallel}D}}\Big)  \Big]e^{-(\sqrt{\frac{1+f(b)}{2}}  -1 )\sqrt{\frac{L^2}{\tau_{s}D}}},
\end{eqnarray}
\end{widetext}
with $f(b)=\sqrt{1+b^2}$ and $b=B/B_{sup}$ the reduced magnetic field, where $B_{sup}$ is the characteristic field for spin polarization suppression\cite{js1985} $B_{sup} =(\gamma_c B \tau_{s})^{-1}$.

By inspecting Eqs. (\ref{Eq4})-(\ref{Eq5}), it is apparent that the solution of $V_{nl}(B,\beta)$ for arbitrary $\beta$ can be obtained by simply rescaling $V_{nl}(B_\perp)$ by a factor $\sin^2(\beta)$ and then adding a field-independent offset, $\cos^2(\beta)$. This result is not unexpected given that the precession dynamics of the spin component perpendicular to the field should be independent of $\beta$; only the magnitude of the perpendicular component and its projection along the detector magnetization varies.  As such, from a spin precession measurement performed at a known $\beta$, typically $\beta=90^{\circ}$, one can extract $\tau_s$ and $D_s$ by means of a least square fit of the measured lineshapes.

Conversely, Eqs. (\ref{Eq4})-(\ref{Eq5}) also show that one can obtain $V_{nl}(B_\perp)$ from the experimental results at any $\beta\neq0$. Because of the exponential factor in Eq. (\ref{Eq5}), $V_{nl}(B,\beta)\rightarrow \cos^2(\beta)$ for large enough $B$. Therefore, $V_{nl}(B_\perp)$ can be extracted by subtracting from $V_{nl}(B,\beta)$ its asymptotic value and normalizing the result to the value at $B=0$ (note that, by definition, $V_{nl}(B_\perp=0)=1$). This operation does not require knowing $\beta$. As discussed in the following sections, this scaling of the spin precession response at different $\beta$ applied to arbitrary $\zeta$ provides a straightforward way to characterize the level of anisotropy, as the universal scaling observed in the isotropic case, $\zeta=1$, breaks down when $\zeta \neq 1$.


\begin{figure}[t]
\includegraphics[width=1.1\linewidth]{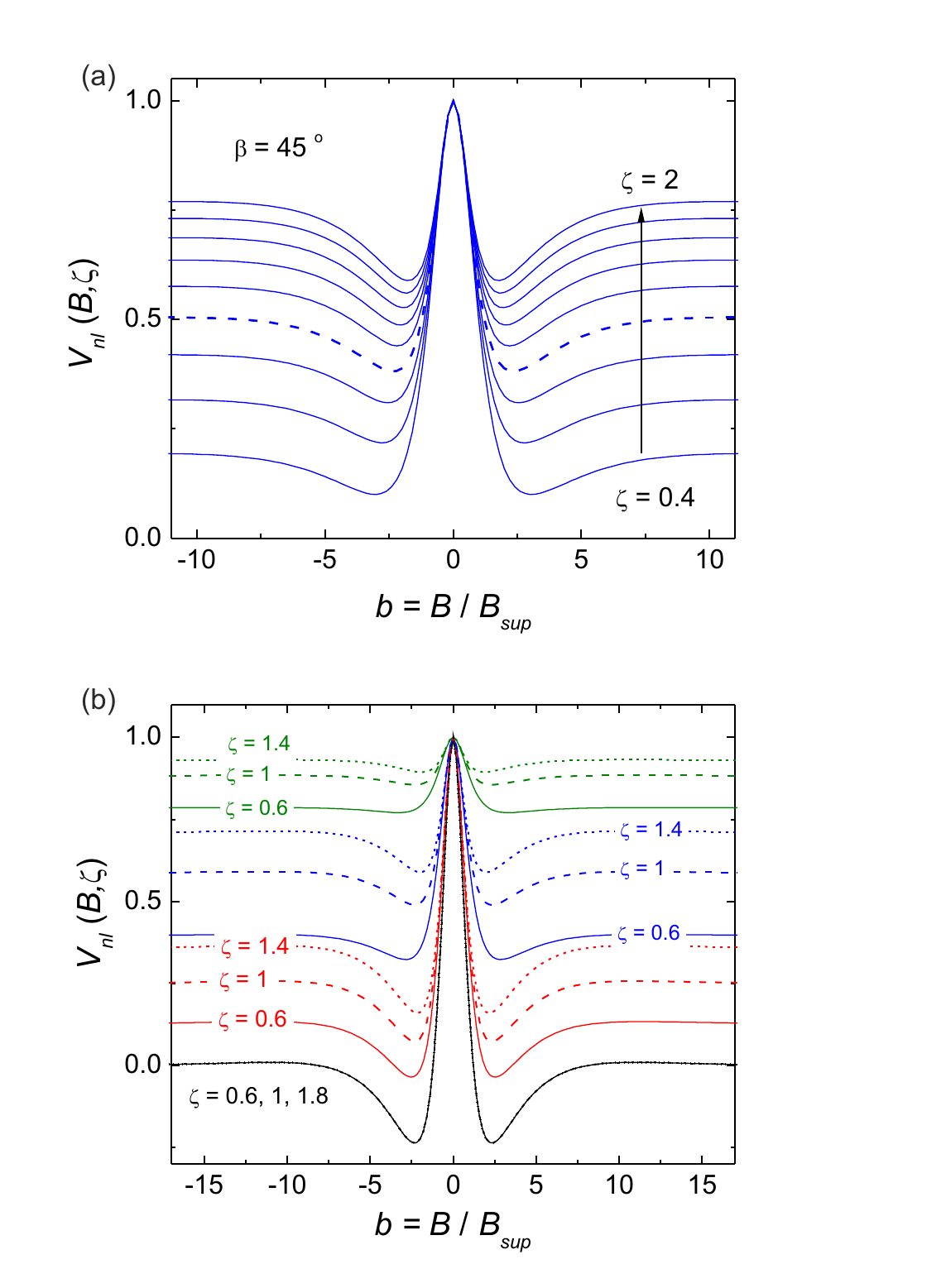}
\caption{Anisotropic spin precession. Spin precession (Hanle) lineshapes $V_{nl}(B,\zeta)$ are obtained numerically using a generalized model for spin diffusion in an anisotropic medium and then normalized to their value at zero magnetic field. (a) Calculations at fixed $\beta=45^{\circ}$ for anisotropy ratios $\zeta$ between 0.4 and 2 in steps of 0.2. The dashed line corresponds to the isotropic case, $\zeta =1$. As $\zeta$ increases, the overall spin signal and, in particular, its asymptotic value increase. (b) Spin precession lineshapes for $\beta=$ $90^{\circ}$ (black), $60^{\circ}$ (red), $40^{\circ}$ (blue), and $20^{\circ}$ (green). In each case, we show the numerical results for $\zeta=$ 0.6 (solid line), 1 (dashed line) and 1.4 (dotted line). In the calculations we take $\tau_{s\parallel}$ to be constant, independently of the anisotropy, and only $\tau_{s\perp}$ varies. Because at $\beta=$ $90^{\circ}$ the spins precess within the plane of the sample, the spin dynamics is determined by $\tau_{s\parallel}$ only, and the results are independent of $\zeta$.}\label{Fig2}
\end{figure}


\section{Anisotropic spin precession with arbitrary $\beta$} \label{aniso}

In order to investigate the effect of the anisotropy, $\zeta\neq1$, on the spin precession lineshapes, we solve Eq. (\ref{Eq1}) numerically. We apply a backward time, centered space method and introduce Neumann boundary conditions. We consider that the magnetization of the injector electrode, and thus the spin polarization of the current at $x=0$, is fixed along the $y$-axis, with null components along $x$ and $z$.  Before carrying out the numerical calculations for the anisotropic case, we verified that the analytical results of the isotropic case were reproduced exactly.

We now discuss spin precession in an anisotropic system by evaluating Eq. (\ref{Eq1}) using typical spin properties of graphene, namely, $D_s=0.02$ m$^2$s$^{-1}$ and $\tau_{s \parallel}=1$ ns, which result in an in-plane spin relaxation length $\lambda_{s \parallel}=\sqrt{D_s\tau_{s \parallel}} \sim 4.5$ $\mu$m. The length of the channel between injector and detector is chosen to be $L=2.12\lambda_{s \parallel}$. The boundary conditions at $x=\pm\infty$ are fulfilled in practice by taking a system longer than $5\lambda_s$ away from the injection point. Note, however, that the general conclusions are independent of the particular choice of parameters and can be applied to higher or lower mobility graphene or other 2DMs.

\begin{figure}[b]
\includegraphics[width=1.1\linewidth]{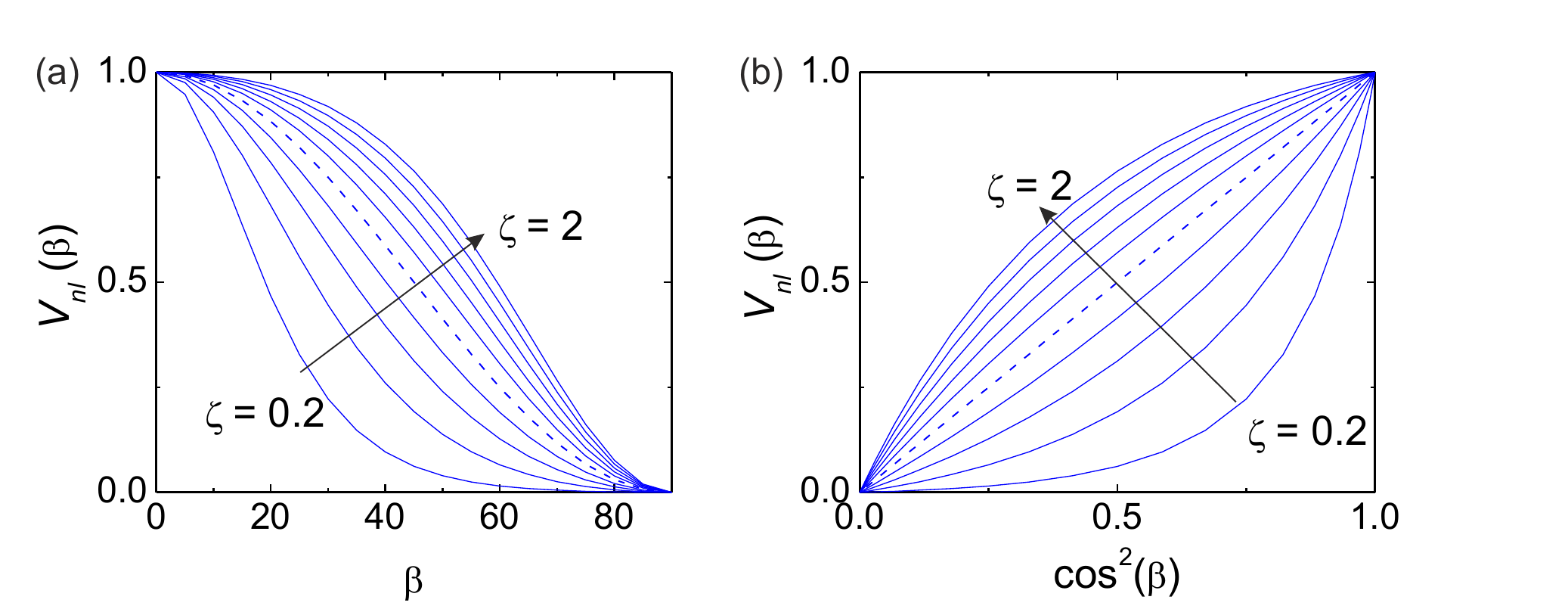}
\caption{Asymptotic value of the spin signal $V^{\infty}_{nl}$, after normalizing at $B=0$, as a function of $\beta$ (a) and $\cos^2 \beta$ (b). The calculations are for $\zeta$ between 0.2 and 2 in steps of 0.2 using the analytical formula represented by Eqs. (\ref{Eq6}) and (\ref{Eq7}). The dashed line corresponds to the isotropic case, $\zeta =1$. In panel (b), the $\zeta =1$ curve appears as a straight line. Therefore, deviations from the straight line and the curvature of the results readily indicate the degree of anisotropy and whether $\zeta < 1$ or $\zeta > 1$. We propose direct fitting of experimental results to Eqs. (\ref{Eq6}) and (\ref{Eq7}) to determine the value of $\zeta$.}\label{Fig3}
\end{figure}

Figure \ref{Fig2} outlines the main signatures of the obtained spin precession response $V_{nl}$ as a function of $b$, $\beta$ and the anisotropy ratio $\zeta$. Given that the spin relaxation time is no longer a scalar, we define $b=B/B_{sup}$ where now $B_{sup}=(\gamma_c B \tau_{s \parallel})^{-1}$. Figures \ref{Fig2}(a) and (b) show $V_{nl}$ for the specified values of $\zeta$ as a function of $b$ for a fixed $\beta=45^\circ$ (a) and as a function of $b$ for four different $\beta$ (b). We readily observe that $V_{nl}$ is strongly dependent on $\zeta$. In particular, its asymptotic value $V^{\infty}_{nl}$ at large $b$ (when the spin precession is suppressed due to diffusive broadening) increases monotonically with its magnitude. Figure \ref{Fig3}(a) shows $V^{\infty}_{nl}$ versus $\beta$. The curves above the one corresponding to $\zeta=1$ are for $\zeta>1$ while the curves below it are for $\zeta <1$. This demonstrates that the anisotropy can be quickly visualized from the experimental results in such a representation.


Furthermore, the response $V^{\infty}_{nl}(\beta,\zeta)$ can be determined analytically as an asymptotic solution of the Bloch equations. When the precessional motion is completely suppressed, the components of the spin density that are perpendicular to the magnetic field direction cancel out, $s_x=0$ and $s_{B_\perp}=0$, and the equation for $s_{B_\parallel}$ decouples. The normalized contribution of $s_{B_\parallel}$ to the nonlocal voltage at the detector electrode then becomes
\begin{eqnarray}\label{Eq6}
V_{nl}^{\infty}(\beta,\zeta)=\sqrt{\frac{\tau_{s_B}}{\tau_{s\parallel}}} e^{-\sqrt{\frac{L^2}{\tau_{s\parallel}D_s}}(\sqrt{\frac{\tau_{s\parallel}}{\tau_{s_B}}}-1)} \cos^2(\beta),
\end{eqnarray}
with
\begin{eqnarray}\label{Eq7}
\frac{\tau_{s_B}}{\tau_{s\parallel}}=\big(\cos^2(\beta)+\frac{1}{\zeta}\sin^2(\beta)\big)^{-1}.
\end{eqnarray}

When $\zeta=1$, it follows from Eq. (\ref{Eq7}) that $\frac{\tau_{s_B}}{\tau_{s\parallel}}=1$, which leads to $V^{\infty}_{nl}(\beta,\zeta=1)=\cos^2(\beta)$, as derived in the previous section and shown by Eq. (\ref{Eq4}). Equations (\ref{Eq6}) and (\ref{Eq7}) form the basis for the method introduced in Ref. \onlinecite{BR2016} to determine $\zeta$ as a single fitting parameter. Plotting $V^{\infty}_{nl}(\beta,\zeta=1)$ versus $\cos^2(\beta)$ results in a straight line. According to Eq. (\ref{Eq6}), the lineshapes $V^{\infty}_{nl}(\beta,\zeta>1)$ lie above this straight line while the lineshapes $V^{\infty}_{nl}(\beta,\zeta<1)$ lie below it (see Fig. \ref{Fig3}(b)).


\begin{figure}[t]
\includegraphics[width=1.1\linewidth]{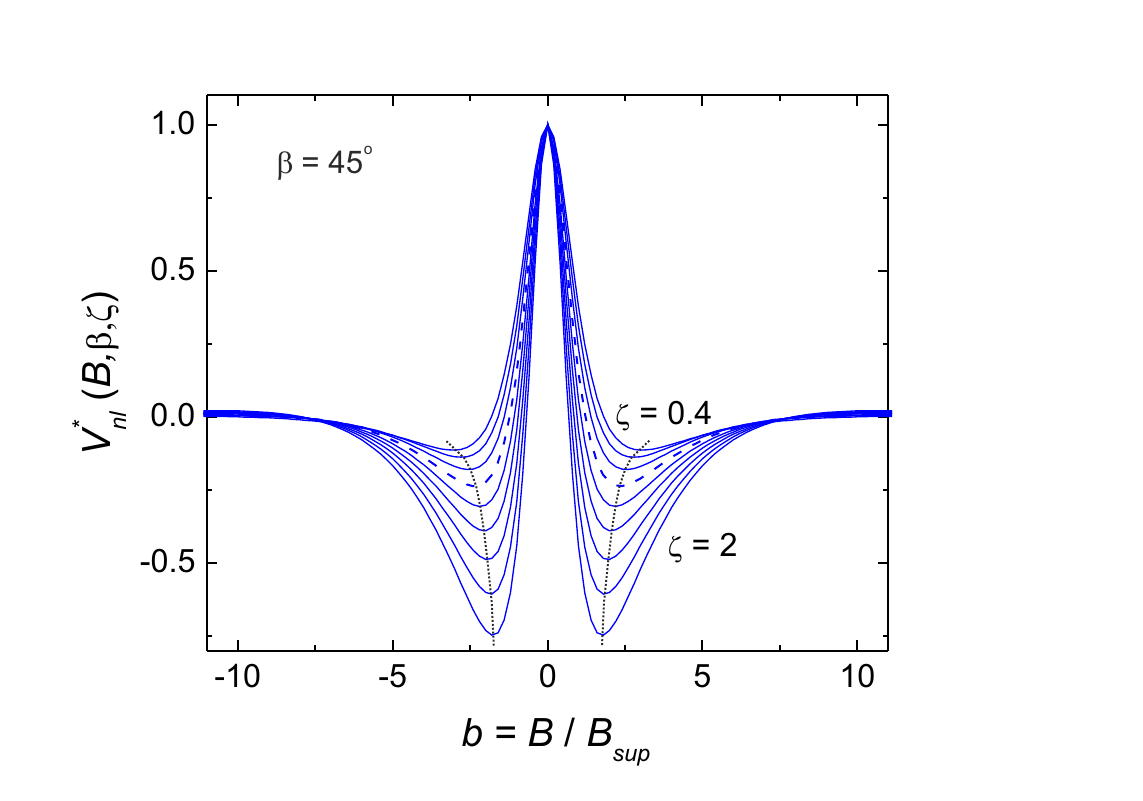}
\caption{Scaled spin precession (Hanle) lineshapes $V^*_{nl}(B,\zeta)$, obtained numerically. We have subtracted the asymptotic value of each curve in Fig. \ref{Fig2}(a) and normalized the result to its value at $B=0$.  Calculations are carried out at fixed $\beta=45^{\circ}$ for anisotropy ratios $\zeta$ between 0.4 and 2 in steps of 0.2. The dotted lines mark the position of the minima. We observe that for larger $\zeta$ the minima appear at smaller $B$ with increasing magnitude. This is a consequence of the increasing $\tau_{s\perp}$.}\label{Fig4}
\end{figure}


Further inspection of Figs. \ref{Fig2}(a) and (b) shows that the oscillatory fingerprint of the spin precession at low magnetic field magnitudes also reflects the degree of anisotropy for spin transport in the channel. Specifically, as $\zeta$ increases the minima in the Hanle response become sharper and develop at lower $B$. This is more clearly seen in Fig. \ref{Fig4}, where the results of Fig. \ref{Fig2}(a) have been plotted to highlight the oscillatory profile due to spin precession. As proposed at the end of Section \ref{iso} to indicate the universal scaling in the isotropic case, we have subtracted  $V^{\infty}_{nl}$ from each curve and normalized the result to its value at $B=0$. We therefore define the rescaled $V^*_{nl}(B,\beta,\zeta)$ as

\begin{eqnarray}
V^*_{nl}(B,\beta,\zeta) = \frac{[V_{nl}(B,\beta,\zeta)-V^{\infty}_{nl}(\beta,\zeta)]}{[V_{nl}(0,\beta,\zeta)- V^{\infty}_{nl}(\beta,\zeta)]}.
\end{eqnarray}

That the minima have a larger relative magnitude when increasing $\zeta$ is a consequence of a larger $\tau_{s\perp}$. Because of the oblique magnetic field, the spins precess out of plane and therefore may relax at a different rate than when they remain in plane (at $B=0$). Taking $\zeta=1$ as a reference, the minima for $\zeta<1$ ($\zeta>1$) will decrease (increase) in amplitude because of the enhanced (suppressed) relaxation rate of the spins oriented perpendicular to the channel plane.

The position of the minima, which occurs for a collective $\pi$ spin precession, is also determined by a change in the effective $\tau_s$. The probability that a spin contributes a precession angle $\phi$ is determined by the probability $P(t)$ that a spin injected at the source electrode reaches the detector in a diffusion time $t$, such that $t=\phi/\omega_L(B)$, with $\omega_L(B)$ the Larmor frequency. Here $P(t)$ is determined by the product of the diffusion-time distribution function and the probability that the spin has \emph{not} flipped during $t$. The latter is proportional to $\exp (-t/\tau_s)$, resulting in a suppressed probability at long $t$ or, equivalently, at large $\phi$. Such suppression is more significant for short $\tau_s$, which implies that the collective spin precession angle for a given $B$ will increase with $\tau_s$. This explains why the minima in Fig. \ref{Fig4} develop at lower $B$ as $\zeta$ increases; at larger $\zeta$, that is, larger effective $\tau_s$, a lower $B$ is required to reach a collective $\pi$ rotation.

Similar arguments can be drawn when discussing the response for fixed $\zeta \neq 1$ and variable $\beta$. In that case, the spin component perpendicular to the plane, and therefore the amplitude of the minima, will be determined by $\beta$. This demonstrates the break-down of the universal scaling (that is expected in the isotropic case, $\zeta=1$) when $\zeta \neq 1$. Note that the relative change of the rescaled signal is larger for smaller $\beta$, even though the oscillatory component of $V_{nl}(B,\beta)$ and the absolute change of the non-precessing spin component decrease [see Fig. \ref{Fig2}(b)]. The reason is that the spin component that is perpendicular to the field precesses within a plane that is closer to the normal of the substrate, and therefore its dynamics becomes more sensitive to $\tau_{s\perp}$. This suggests that the best experimental configuration to detect changes in the Hanle lineshape involves applying the magnetic field \emph{along} the graphene channel and \emph{perpendicular} to the easy magnetic axis of the ferromagnetic electrodes. In this case, the spins will precess within a plane that will contain the normal of the substrate and the influence of $\tau_{s\perp}$ would be maximized. The disadvantage of this configuration is that the magnetization of the electrodes would significantly tilt under the influence of the magnetic field, complicating the analysis of the results. We will discuss the analysis of experiments in this configuration elsewhere.

\begin{center}
\begin{figure}[h]
\includegraphics[width=0.85\linewidth]{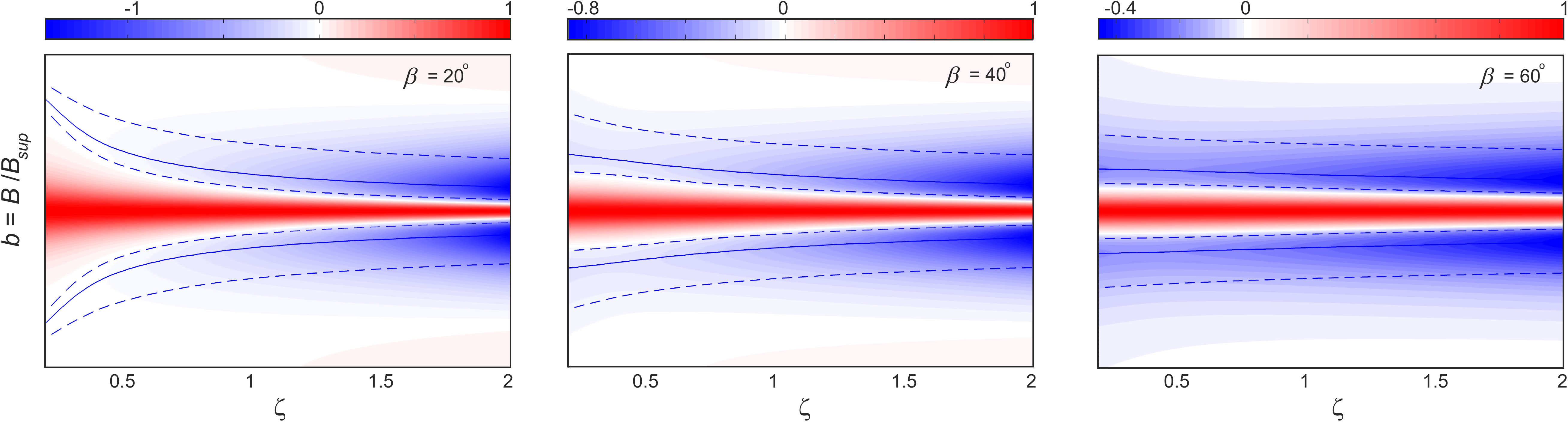}
\caption{Scaled spin precession (Hanle) lineshapes $V^*_{nl}(B,\zeta)$ obtained numerically (color scale) at fixed $\beta=20^{\circ}$, $40^{\circ}$ and $60^{\circ}$ for anisotropy ratios $\zeta$ between 0.2 and 2. The solid (blue) lines mark the position of the minima, and the dashed lines mark half the minimum value.}\label{Fig5}
\end{figure}
\end{center}

Figure \ref{Fig5} summarizes the change of the scaled response $V^*_{nl}(B,\beta,\zeta)$ (color) for $\beta=20^{\circ}$, $40^{\circ}$ and $60^{\circ}$ and $\zeta$ between 0.2 and 2. The solid lines show the position of the minima as a function of $\zeta$ (as in Fig. \ref{Fig4}), whereas the dashed lines mark the value of $b$ at which $V^*_{nl}(B,\beta,\zeta)$ reaches half its minimum value. Independently of $\beta$, the magnitude of the minima is observed to increase with $\zeta$, while their position shifts to lower $b$. As expected from the above discussion, the relative change is larger at smaller $\beta$; for small enough $\beta$ and $\zeta>1$, the magnitude of the minima in $V^*_{nl}(B,\beta,\zeta)$ can even be larger than the maximum at zero magnetic field (see results for $\beta=20^{\circ}$).

\onecolumngrid
\twocolumngrid

\subsection{Contact-induced spin relaxation}

When investigating the origin of the spin relaxation in an actual device, as represented in Fig. \ref{Fig1a}, the influence of the contacts has to be evaluated. In general, the overall spin relaxation rate will be determined by all the relaxation channels available, and spin absorption processes by the contacting electrodes could play an important role. Spin absorption effects in a ferromagnet are not necessarily isotropic; the different absorption of longitudinal and transverse spins must be accounted for by the spin mixing conductance \cite{PhysRevB.89.081308.2014}, which could significantly complicate the analysis. From this point of view, and in order to reduce the number of unknown parameters, it is convenient to suppress the influence of the electrodes, and thus the devices should be optimized to achieve this goal.

The effect of spin absorption on the spin relaxation has been a topic of recent debate, in particular when dealing with graphene-based devices (see, for example, Refs. \onlinecite{PhysRevB.67.052409.2003, PhysRevB.86.235408.2012,PhysRevB.89.245436.2014,PhysRevB.90.165403.2014,PhysRevB.91.241407.2015,arXiv.1512.02255.2015,PhysRevB.94.094431.2016}).
It has been demonstrated, both theoretically and experimentally, that the influence of the electrodes can be minimized by having \textit{i}) large contact resistances $R_\mathrm{c}$ and \textit{ii}) a separation between injector and detector $L$ that is substantially larger than $\lambda_s$. The argument is simple: the contact resistance reduces the flow of spins between the metallic electrodes and the channel, whereas for large $L$ most of the spin relaxation and diffusion takes place in the channel without interference of the electrodes.
A relatively large $L>2 \lambda_s$ is a requirement for the spin anisotropy measurements, so as to achieve the full spin precession response, including dephasing, at the moderate magnetic fields for which the magnetization of the electrodes stay in plane \cite{BR2016}. Large contact resistances must also be introduced to achieve an efficient spin injection.

Considering an effective contact-induced spin relaxation rate $\Gamma_\mathrm{c}$, the measured in-plane and perpendicular spin lifetimes would be given by $(\Gamma_{\parallel}+\Gamma_\mathrm{c})^{-1}$ and $(\Gamma_{\perp}+\Gamma_\mathrm{c})^{-1}$, respectively, where $\Gamma_{\parallel}$ and $\Gamma_{\perp}$ are the channel in-plane and perpendicular spin relaxation rates. When considering the influence of the contacts, the fitted anisotropy ratio $\zeta$ is then

\begin{eqnarray}\label{EqS0}
\zeta = \frac{\Gamma_{\parallel}+\Gamma_\mathrm{c}}{\Gamma_{\perp}+\Gamma_\mathrm{c}} = \zeta^\mathrm{Ch} \frac{1+\kappa}{1 + \zeta^\mathrm{Ch} \kappa} \approx \zeta^\mathrm{Ch}[1+\kappa(1-\zeta^\mathrm{Ch})],
\end{eqnarray}
\\
where $\zeta^\mathrm{Ch}= \Gamma_{\parallel}/\Gamma_{\perp}$ is the actual anisotropy ratio of the channel, and $\kappa=\Gamma_\mathrm{c}/\Gamma_{\parallel}$.

Using the above expression, it becomes evident that $\zeta$ is rather insensitive to $\Gamma_\mathrm{c}$. For typical values of $\kappa$ obtained in experiments\cite{BR2016}, $\kappa \sim 0.1$, when $\zeta^\mathrm{Ch} = 0.9$ we obtain $\zeta \approx 0.908$, which underestimates $\zeta^\mathrm{Ch}$ by less than 1\%. For larger anisotropies, $\zeta$ further deviates from $\zeta^\mathrm{Ch}$. However, even when $\zeta^\mathrm{Ch} = 0.5$, $\zeta \approx 0.52$ and the difference between them is still below 5\%.

\section{Experimental results} \label{Exp}

In the previous sections we demonstrated that the spin relaxation anisotropy can be obtained by two complementary approaches, either by focusing on the dephased nonlocal signal or by analyzing the spin precession lineshape. The first approach was the subject of a prior experimental article, in which the spin relaxation of graphene on SiO$_2$ was found to be isotropic \cite{BR2016}. Here we demonstrate the implementation of the second approach. We present experimental results for two graphene devices, D1 and D2, with spin lifetimes that differ by one order of magnitude. Consistent with previous work, we find no signature of anisotropic spin relaxation.

The graphene flakes for the devices were obtained by mechanically exfoliating highly-oriented pyrolytic graphite (SPI Supplies) onto a $p$-doped Si substrate, which is used as a backgate, covered with 440 nm of SiO$_2$. We chose long uniform graphene flakes, with a channel length of 10 $\mu$m and widths of 1.5 and 0.3 $\mu$m for devices D1 and D2, respectively. To define the electrodes, we used a single electron-beam (e-beam) lithography step and shadow evaporation to minimize processing contamination \cite{J.Vac.Sci.Technol.B.30.04E105.2012}. We first deposited the outer electrodes of Ti(5 nm)/Pd(10 nm) by angle deposition. Selecting an angle of $\pm 45^{\circ}$ from the normal to the substrate ensured that no image of the lithographically-produced lines reserved for the Co electrodes were deposited during the evaporation of Ti/Pd; indeed, Ti and Pd deposit onto the sidewalls of the lithography mask and were later on removed by lift-off. An amorphous carbon (aC) interface was created between all contacts and the graphene flake by e-beam induced deposition prior to the fabrication of the contacts \cite{ApplPhysLett.103.112401.2013,NanoLett.15.4000.2015}. The aC deposition allows us to obtain large spin signals by suppressing the conductivity mismatch, and the contact-induced spin relaxation, and was done by an e-beam overexposure of the contact area, with a dose about 30 times the typical working dose of e-beam resists. All of the measurements shown below were carried out at room temperature using an injector current of $10$ $\mu$A.


\begin{figure}[t]
\includegraphics[width=1.1\linewidth]{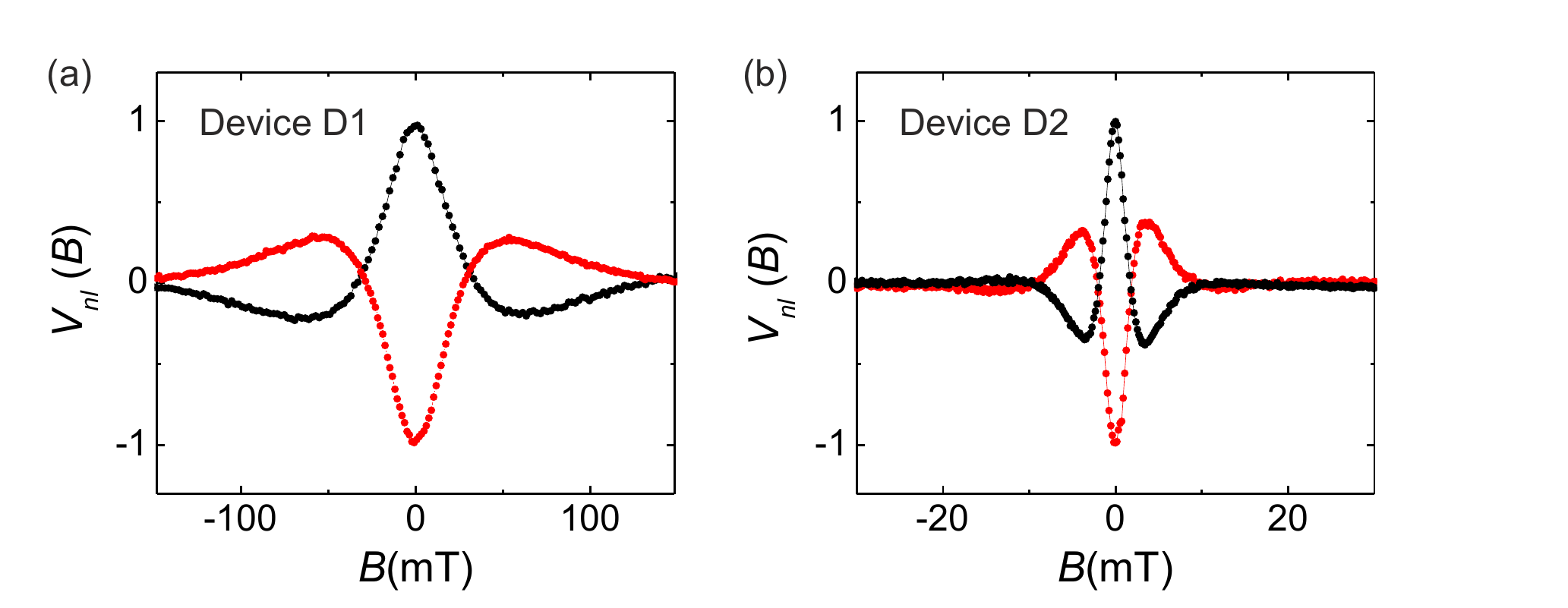}
\caption{Standard spin precession measurements with perpendicular magnetic field $B_{\perp}$ ($\beta=90^{\circ}$) for devices D1 (a) and D2 (b). We show the nonlocal voltage $V_{nl}$, normalized at $B=0$. The black (red) symbols show the measurement for parallel (antiparallel) configurations of the magnetization of the injector/detector ferromagnetic electrodes. }\label{Fig6}
\end{figure}


Figures \ref{Fig6}(a) and \ref{Fig6}(b) show spin precession measurements in the standard configuration with $\beta=90^{\circ}$  and a backgate voltage $V_g = 50$ V for devices D1 and D2, respectively. In both devices the graphene was $n$-doped, with the charge neutrality point at $V_g \sim -15$ V. By fitting the data in Fig. \ref{Fig6} to Eq. (\ref{Eq5}), we obtain $\tau^{D1}_{s\parallel}=0.17$ ns, $D^{D1}_s=0.085$ m$^2$s$^{-1}$, $\tau^{D2}_{s\parallel}=2.5$ ns and $D^{D2}_s=0.0039$ m$^2$s$^{-1}$. We note that device D2 presents a low $D_s$ and somewhat long $\tau_{s\parallel}$ when compared with our typical devices. The anomalous parameters might be related to the very narrow graphene flake and explain the very low magnetic fields that are required for complete dephasing.


\begin{figure}[b]
\includegraphics[width=0.9\linewidth]{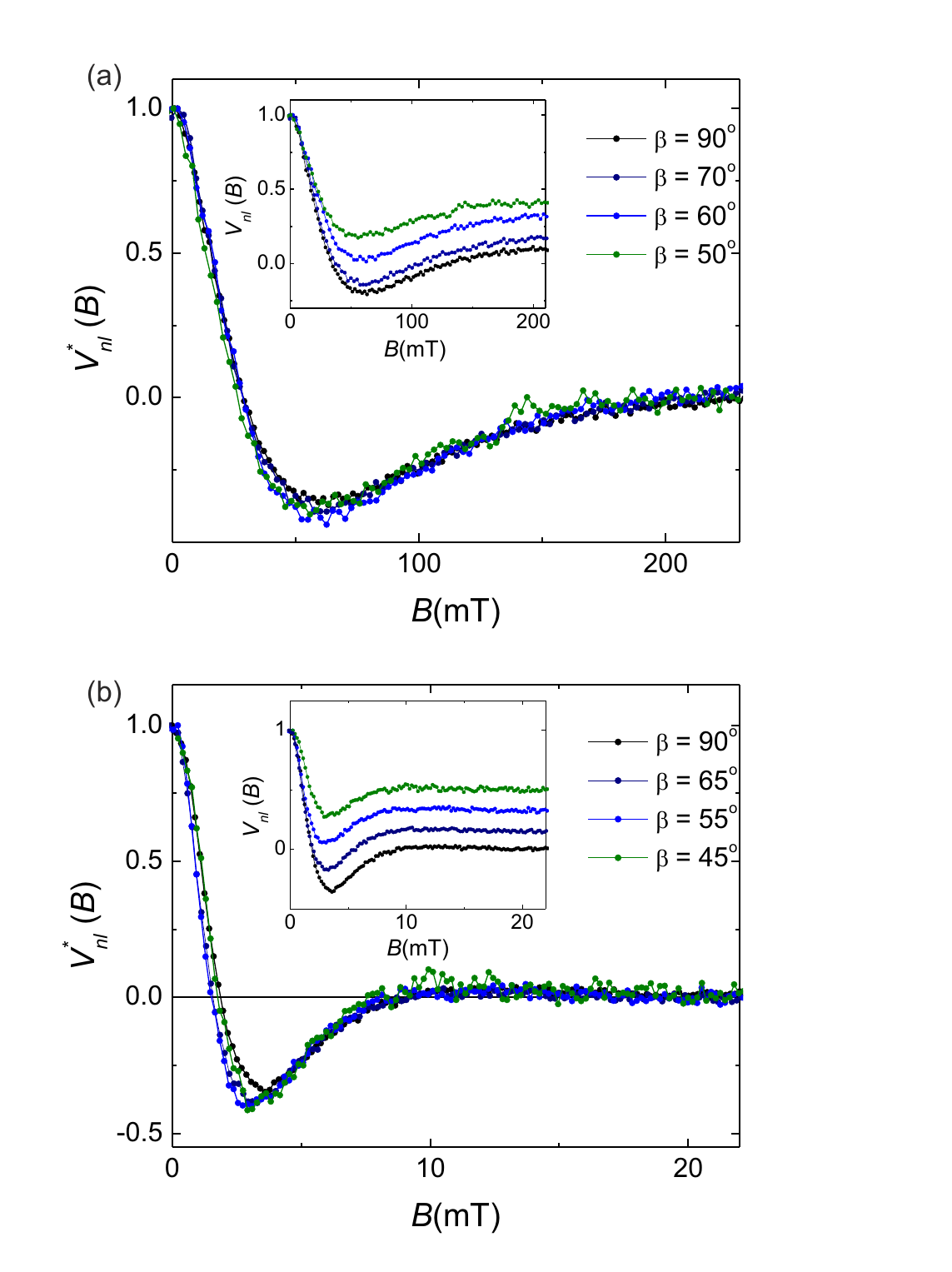}
\caption{Spin precession measurements under oblique magnetic fields for devices D1 (a) and D2 (b). The insets show a representative subset of normalized spin precession curves at the indicated $\beta$. The main panels show the data in the inset after suitable scaling, $V^*_{nl}(B,\zeta)$. All of the curves are observed to collapse into a universal curve, which indicates a low anisotropy and $\zeta \sim1$.}\label{Fig7}
\end{figure}


We now focus on the $\beta$-dependence of the spin precession response. The measurements are shown in the insets of Fig. \ref{Fig7} for device D1 (a) and D2 (b). In all cases, we first prepare the electrode magnetization by applying a large positive $B$ and then acquire $V_{nl}$ while sweeping down $B$. This procedure yields a reproducible configuration of the electrodes in which their magnetizations are uniform and parallel to each other.
In the main panels of Figs. \ref{Fig7}(a) and \ref{Fig7}(b), we present the same results after scaling, $V^*_{nl} (B)$, following the procedure discussed in Sections \ref{iso} and \ref{aniso}. We observe that the measurements at different $\beta$ collapse to a universal curve, which suggests a small anisotropy and $\zeta \sim 1$. There are small deviations about the minima (more clearly seen in device D2). However, these deviations are within the precision of our measurements, which therefore defines the accuracy in the determination of $\zeta$.

\begin{figure}[t]
\includegraphics[width=0.9\linewidth]{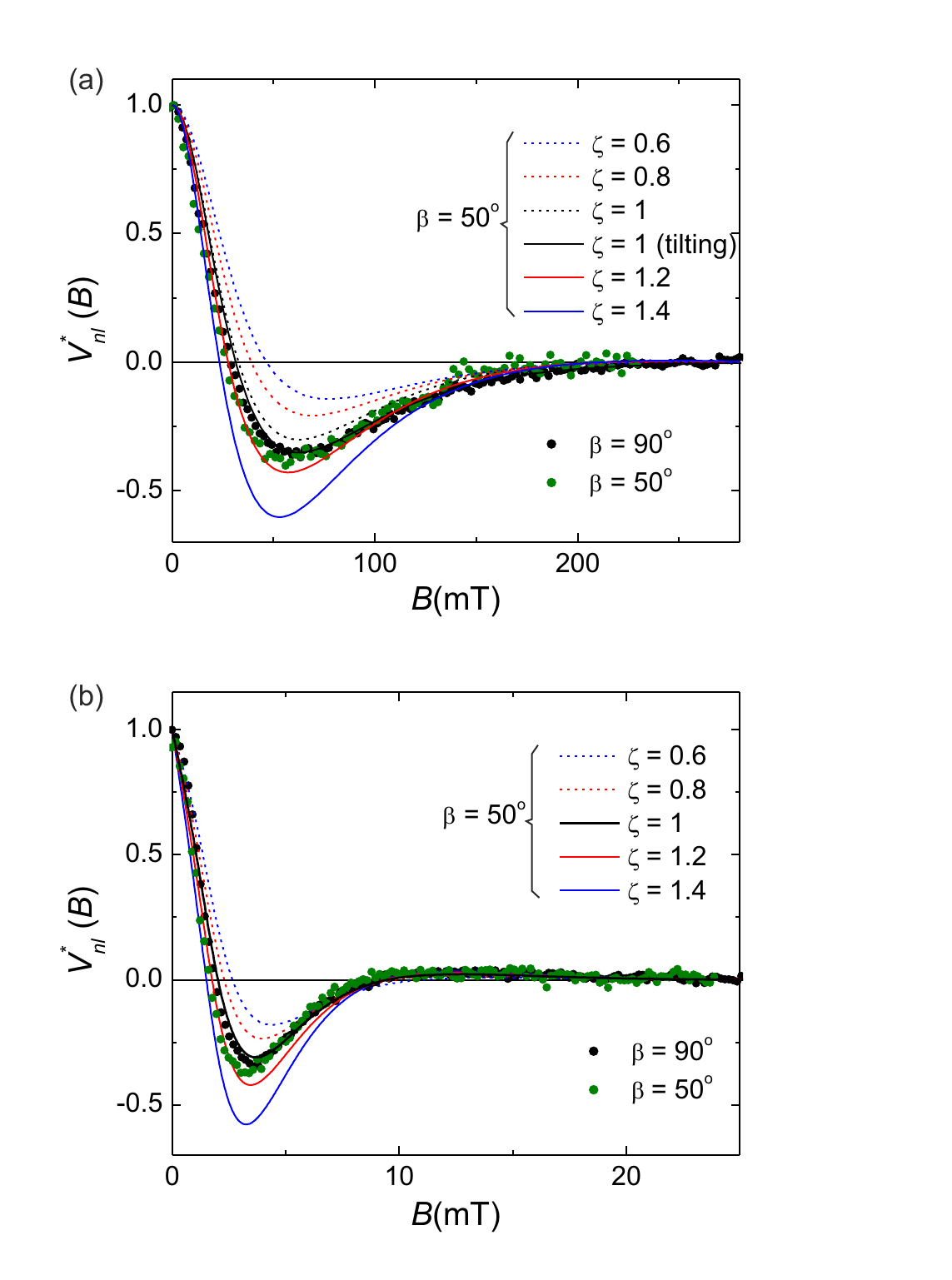}
\caption{Determination of the spin relaxation anisotropy by comparison with numerical simulations for devices D1 (a) and D2 (b). We show experimental data $V^*_{nl}(B,\zeta)$ for the two devices at $\beta=90^{\circ}$ and $\beta= 50 ^{\circ}$ and numerical results for the same angles for $\zeta$ between 0.6 and 1.4. The calculations for $\beta=90^{\circ}$ and $\beta= 50 ^{\circ}$ with $\zeta=1$ coincide. However, the more $\zeta$ deviates from 1, the larger is the difference in the scaled curves, as seen in Fig. \ref{Fig4}.}\label{Fig8}
\end{figure}

Figure \ref{Fig8} compares the experimental results for $\beta=90^{\circ}$ and $50^{\circ}$ (circles) with the solutions (solid and dash lines) of the scaled anisotropic spin precession signal $V_{nl}^*$, which allow us to determine upper and lower limits for the values of $\zeta$ in our samples. The modelling was performed for $\beta=90^{\circ}$ and $50^{\circ}$ to match the experiments. We used the quoted $\zeta$, namely 0.6, 0.8, 1, 1.2 and 1.4, and the parameters $\tau^{D1, D2}_{s\parallel}$ and $D^{D1,D2}_s$ for devices D1 (a) and D2 (b), as obtained above with the standard spin precession measurements. As expected, the model shows a perfect scaling for the case $\zeta=1$ and an increase (decrease) in the damping of the Hanle oscillation when the magnetic field is tilted for $\zeta<1$ ($\zeta>1$).

The degree of anisotropy can be determined by identifying the corresponding $\zeta$ values of the theoretical curves that enclose the experimental results. For this, it is necessary to consider the noise level and any possible systematic deviations in the measurements. The experimental results for device D1 [Fig. \ref{Fig8}(a)] are seen to be enclosed by the curves corresponding to $\zeta =1.2$ (solid red line) and $\zeta = 1$ (dashed black line). However, note that for this device, a change in the electrode magnetization orientation of about $5^{\circ}$ is expected at $\sim$200 mT, which introduces a systematic deviation between the scaled experimental curves and the theoretical curves, given that the latter do not take such magnetization tilting into account. Because the tilting adds a positive signal that increases with $B$, the $\zeta$ obtained after the scaling appears to be larger than it actually is by $\Delta \zeta \sim 0.1$. This can be verified by introducing the tilting in the calculations for $\beta=90^{\circ}$, where we find that the fitting with the experiment is excellent (black solid line). Similar deviations have been observed when using $V^{\infty}_{nl}$ to determine $\zeta$ when the tilting is not considered \cite{BR2016}. By noting that the change in the minima between $\zeta =1.2$ and $\zeta = 1$ is similar to that between $\zeta =1$ and $\zeta = 0.8$ [Fig. \ref{Fig8}(a)], and that it is also comparable to the noise level, we can confidently state that $0.8 <\zeta < 1.2$ or, when taking into account the tilting of the magnetization of detector and injector electrodes, $\zeta = 1\pm 0.1$.

The same analysis can be carried out for device D2. In this case, complete dephasing is obtained at $B$ just above 10 mT. At such small $B$, the tilting of the magnetization of the electrodes is below $1^{\circ}$ and can be disregarded. This is reflected in the constant $V_{nl}$ over a magnetic field range well beyond 20 mT, and in the agreement with the numerical results for $\beta =90^{\circ}$, which is excellent. The boundaries for $\zeta$ can thus be directly found by comparing the noise level of the measurements with the changes of the Hanle modulation associated with $\zeta$. Despite the differences in the characteristic spin relaxation time and diffusion constant with device D1, we find that the scaling yields the same result. In this case, we conclude that $\zeta = 1\pm0.2$. 


\section{Discussion and Conclusions} \label{Concl}

We have proposed a model based on the anisotropic Bloch equations, and have discussed the evolution of the signal that would be detected in a specifically designed nonlocal spin injection and detection configuration under oblique magnetic field $\vec{B}$. We have demonstrated that the asymptotic spin signal after dephasing at large enough $B$, and the oscillatory response associated with the spin precession, are both strongly dependent on the spin relaxation anisotropy ratio $\zeta$, which led us to propose two complementary methods to determine $\zeta$. In the first method, $\zeta$ is obtained from a direct fit to an analytical expression of the dephased spin signal as a function of the orientation of the magnetic field. In the second method, we propose a scaling of the spin signal for two or more orientations of the magnetic field. In the case of an isotropic channel, such a scaling would result in a universal curve, as we have demonstrated both analytically and numerically. We have implemented these ideas using nonlocal devices based on graphene with very different spin lifetimes and found that the relaxation is isotropic within the uncertainty of our measurements.

The main advantage of the second method is that it allows us to quickly visualize any deviations from the isotropic case, and that it does not require an accurate determination of the magnetic field orientation. The main disadvantage is that $\zeta$ is roughly estimated following a comparison with the model, which is limited by the noise level and the fact that the tilting angle of the electrode magnetization is magnetic field dependent. Improved determination of $\zeta$ must therefore be accompanied by an improved signal-to-noise ratio and a precise determination of the tilting. This is less critical in the first method, where the direct fitting at fixed magnetic field allows for a more accurate determination of $\zeta$ . Despite this, quick inspection of the scaled results in our graphene devices enabled us to set lower and upper limits for $\zeta$ of about 0.8 and 1.2, respectively. Indeed, our results indicate that the expected anisotropy that would follow from (Rashba) in-plane spin-orbit fields in ultra clean graphene, with $\zeta$ in the range of 0.5 to 0.6, should still be readily resolved, making the scaling a powerful and versatile technique for spin anisotropy characterization.

Finally, our simple theoretical analysis opens the way for studies involving oblique spin precession, which can be extremely rich. In future spin devices using two-dimensional systems, it will likely be possible to tune the spin-orbit interaction by gating, which will therefore tune the anisotropy of the spin relaxation. If the device is implemented with a fixed magnetic field, its output will be modulated by the anisotropy and thus by the gate. Oblique spin precession will also allow for fundamental studies of spin-orbit control by impurities or by proximity effects between different materials and substrates, which are fundamental, for example, to understand and manipulate spin relaxation in graphene \cite{NatNanotechnol.9.324.2014,JPhysD.47.094011.2014,2DMaterials.2.030202.2015, PhysRevLett.112.116602.2014, NatPhysics.10.857.2014}.

\vspace{5mm}
\section*{Acknowledgments}

We thank D. Torres for his help in designing Fig. 1. This research was partially supported by the European Research Council under Grant Agreement No. 308023 SPINBOUND, by the European Union's Horizon 2020 research and innovation programme under grant agreement No. 696656, by the Spanish Ministry of Economy and Competitiveness, MINECO (under Contracts No. MAT2013-46785-P, No MAT2016-75952-R, No. FIS2015-67767-P and Severo Ochoa No. SEV-2013-0295), and by the Secretariat for Universities and Research, Knowledge Department of the Generalitat de Catalunya.
M.V.C. and J.F.S. acknowledge support from the Ram\'on y Cajal and Juan de la Cierva programs, respectively.


\end{document}